\documentclass[twocolumn,preprintnumbers]{revtex4}

\usepackage[dvipdfmx]{graphicx,color}
\usepackage{epsf}
\usepackage{amsmath,amssymb}
\usepackage{bm}
\usepackage{color}
\usepackage{textcomp}

\newcommand{\hatrho}{\hat\rho^{\textrm{(1)}}_{\Psi^{(2)}}}
\newcommand{\hatrhotwo}{\hat\rho^{\textrm{(2)}}_{\Psi^{(2)}}}
\newcommand{\hatrhoG}{\hat\rho^{(G)}_{\Phi_G}}
\newcommand{\frho}{\rho^{\textrm{(1)}}_{\Psi^{(2)}}}
\newcommand{\frhocl}{\rho^{\textrm{(1)},\textrm{cl}}_{\Psi^{(2)}}}
\newcommand{\hatrhocl}{\hat\rho^{\textrm{(1),cl}}_{\Psi^{(2)}}}
\newcommand{\rhow}{\rho^{\textrm{W}}}

\newcommand{\rhohh}{\rho_{\hbar/2}^\textrm{H}}
\newcommand{\R}{s}
\newcommand{\pa}{{\rm c1}}
\newcommand{\pb}{{\rm c2}}
\newcommand{\rG}{R}
\newcommand{\bs}{b_r}
\newcommand{\SR}{S^{\textrm{R2}}}
\newcommand{\SRcl}{S^{\textrm{R2,cl}}}
\newcommand{\dqdp}{\frac{dqdp}{2\pi\hbar}}
\newcommand{\clapprox}{\stackrel{\textrm{cl}}{\approx}}

\newcommand{\loge}{\textrm{ln}}
\begin{document}
\preprint{}
\title{Quantum decoherence in the entanglement entropy of a composite particle and its relationship to coarse graining in the Husimi function}
\author{Yoshiko Kanada-En'yo}
\affiliation{Department of Physics, Kyoto University, Kyoto 606-8502, Japan}

\begin{abstract}
I investigate quantum decoherence in a one-body density matrix of a composite particle consisting
of two correlated particles. 
Because of a two-body correlation in the composite particle, quantum decoherence occurs 
in the one-body density matrix that has been reduced from the two-body density matrix. 
As the delocalization of the distribution of the composite particle grows, the entanglement entropy increases,
and the system can be well-described by a semi-classical approximation, wherein
the center position of the composite particle can be regarded as a classical coordinate.
I connect the quantum decoherence in the one-body density matrix 
of a composite particle to the coarse graining in a phase space distribution function of a single particle
and associate it with the Husimi function.
\end{abstract}
\maketitle

In recent decades, quantum entanglement has attracted a great deal of interest in various fields.
To estimate correlations in quantum systems, entanglement measures such as
the entanglement entropy (EE)  have been intensively studied 
\cite{grobe94,bennett96,Horodecki01,Calabrese:2004eu,Plenio07,amico08,Horodecki09,Nishioka:2009un,tichy11}. 
Many entanglement measures are defined by reduced density matrices which describe the
structure of the Schmidt decomposition and contain information about entanglements in quantum systems. 
In entangled states,  the EE is produced by quantum decoherence caused by a reduction in the number of
degrees of freedom (DOF).
In my previous papers, I calculated entanglement measures of the one-body density matrix 
in nuclear systems \cite{enyo-ee1,enyo-ee2} and showed that the EE is enhanced by 
the delocalization of the distribution of clusters, which are composite particles of spatially correlated nucleons.
My first aim in the present paper is to understand how quantum decoherence occurs and how the EE is produced 
in the one-body density matrix of correlating particles. 

Quantum decoherence---that is, the quantum entropy---has been also investigated with the coarse graining of 
distribution functions in a phase space. 
The Husimi function \cite{Husimi40,Takahashi86,Nakamura88} is known to 
have finite the Wehrl and R\'enyi-Wehrl entropies \cite{Wehrl:1978zz,Gnutzmann01} because of the coarse graining
by a Gaussian smearing of the Wigner function.
It has been shown 
that the Wehrl and R\'enyi-Wehrl entropies are increased by delocalization of the distributions
in quantum systems \cite{sugita02,sugita03,Kunihiro:2008gv,campos10}. Campos {\it et al.} have discussed a correlation between the EE and the R\'enyi-Wehrl entropy in entangled states \cite{campos10}.
One of the fundamental questions in quantum physics is 
how the quantum decoherence in the reduced density matrix of entangled states
can be connected to the coarse graining of distribution functions. 
My second aim in this paper is to understand the correspondence between the quantum decoherence 
in the one-body density matrix of correlating particles and the coarse graining 
in the Husimi function of a single-particle state.

In this paper, I investigate the EE of the one-body density matrix of 
a two-body system in which two particles are strongly correlated to form 
a composite particle, and I discuss 
how quantum decohence occurs in the reduction of the DOF.
To describe two-body wave functions,
I adopt a cluster wave function in the 
generator coordinate method in nuclear physics \cite{GCM1,brink66}.
Let us consider a system where two 
particles ($\pa$  and $\pb$) with masses $m$ and $um$ 
form a bound state with an attractive inter-particle force.
I assume that the bound state 
is described by the lowest state of a harmonic oscillator (ho) potential
and can be approximately treated as an inert composite particle, where intrinsic 
excitations cost a relatively high amount of energy compared with
the center of mass (cm) motion of  the composite particle. 
In this approximation,  
a total two-body wave function is given as  
\begin{eqnarray}
|\Psi^{(2)}\rangle=
\int d\R F(\R)| \R;b \rangle_1| \R;b_2 \rangle_2, \label{eq:wave-func}\\
\langle r_i|\R;b\rangle_i=
\textrm{e}^{-\frac{1}{2b^2}(r_i-\R)^2}/(b^2\pi)^{1/4},
\end{eqnarray}
where $b_2=b/\sqrt{u}$, and 
$\langle\Psi^{(2)}|\Psi^{(2)}\rangle=1$. $r_1$($r_2$) is the coordinate of \pa(\pb). 
Here, I describe the one-dimensional case, but the present model can also be extended
to the three-dimensional case.
$|\R;b \rangle_1|\R;b_2\rangle_2$ indicates the 
composite particle localized around the mean position $\R$, and $\Psi^{(2)}$ is given by the superposition of different $\R$ states with the weight factor $F(\R)$.
$|\Psi^{(2)}\rangle$ can be expressed by the cm motion and the intrinsic wave functions as
\begin{eqnarray}
|\Psi^{(2)}\rangle&=&|\Phi_G(\rG)\rangle |\phi_{\rm int}(r) \rangle,\\
\langle \rG |\Phi_G(\rG)\rangle&=&\int d\R \frac{F(\R) }{(b_G^2\pi)^{1/4}}\textrm{e}^{-\frac{1}{2b_G^2}(\rG-\R)^2},
\label{eq:cm-motion}
\end{eqnarray}
with the cm coordinate $\rG=u_1r_1+u_2r_2$  ($u_1=1/(u+1)$ and $u_2=u/(u+1)$), 
 the relative coordinate $r=r_1-r_2$, and $b_G=\sqrt{u_1}b$.
Here, $\langle r |\phi_{\rm int}(r) \rangle=
\textrm{e}^{-\frac{1}{2\bs^2}r^2}/{(\bs^2\pi)^{1/4}}$
with $\bs=b/\sqrt{u_2}$ is the lowest intrinsic state for
the ho potential, $U_{\rm ho}(\mu,\bs;r)=-\hbar^2r^2/2\mu \bs^4$, with $\mu=u_2m$.
Thus, general low momentum states of the inert composite particle can be expressed by 
the form \eqref{eq:wave-func}, in which the cm motion $\Phi_G(\rG)$ 
is expressed by the shifted Gaussian expansion as given in Eq.~\eqref{eq:cm-motion}. 

The one-body density matrix $\hatrho$ 
for $\pa$  is defined by the matrix reduced from the many-body density matrix $\hatrhotwo=|\Psi^{(2)}\rangle 
\langle\Psi^{(2)}|$ as $\hatrho=\textrm{Tr}_2[\hatrhotwo]$
and is given as
\begin{eqnarray}
&&\hatrho=
\int d\R d\R' W(\R',\R)  \langle \R';b_2|\R;b_2\rangle
|\R;b\rangle \langle \R';b| ,\label{eq:one-body}\\
&&\frho(q_1,q_1')=\langle q_1|\hatrho|q_1' \rangle\nonumber\\
&&= \int d\R d\R' 
\frac{W(\R',\R)}{(b^2\pi)^{1/2}}\textrm{e}^{-\frac{u}{4b^2}(\R-\R')^2}\nonumber\\
&&\times \textrm{e}^{-\frac{1}{2b^2}(q_1-\R)^2 
-\frac{1}{2b^2}(q_1'-\R')^2},
\end{eqnarray}
where $W(\R',\R)\equiv F^*(\R')F(\R)$ and $\textrm{Tr}\hatrho=1$. 
Note that $\hatrho$ for $u=3$ equals to the one-body density matrix of
an $\alpha$ cluster composed of four nucleons with an equal mass investigated
in previous papers \cite{enyo-ee1,enyo-ee2}.
The Wigner transformation (Wigner function) of $\hatrho$
is 
\begin{eqnarray}
&&\rhow(\hatrho;q_1,p_1)=\int d\eta
\langle q_1+\frac{\eta}{2} |\hatrho| q_1-\frac{\eta}{2} \rangle  \textrm{e}^{-\frac{ip_1\eta}{\hbar}}\nonumber\\
&&=2 \int d\R d\R' 
\frac{W(\R',\R)}{(b^2\pi)^{1/2}}
\textrm{e}^{-\frac{u}{4b^2}(\R-\R')^2}\nonumber\\
&&\times\textrm{e}^{-\frac{1}{2b^2}(q_1-\R)^2 
-\frac{1}{2b^2}(q_1-\R')^2-\frac{b^2}{\hbar^2}
\left\{ p_1-\frac{i\hbar}{2b^2}(\R-\R')\right\}^2}.\label{eq:wigner-one-body}
\end{eqnarray}
The R\'enyi EE of order 2 (R\'enyi-2 EE) and von Neumann EE 
for $\Psi^{(2)}$ with the one-body density matrix $\hatrho$ are given as
\begin{eqnarray}
&&\SR(\hatrho)=
- \loge\left(\textrm{Tr}\left[ \{\hatrho\}^2 \right]\right)\nonumber\\
&&=- \loge\left( \int dq_1dp_1 \{\rhow(\hatrho;q_1,p_1)\}^2   \right),\\
&&S^{\textrm{vN}}(\hatrho)=
- \textrm{Tr}\left[ \hatrho \loge \hatrho \right].
\end{eqnarray}
If $\rhow(\hatrho;q_1,p_1)\ge 0$ is satisfied in the entire phase space, 
I can consider
the phase-space Shannon entropy 
$S^\textrm{Sh}\left(\rho(q,p)\right)=
\int \dqdp \rho(q,p)\loge\rho(q,p)$
for  $\rhow(\hatrho;q_1,p_1)$ as,
\begin{eqnarray}
S^\textrm{W-Sh}\left(\hatrho \right)&= &
S^\textrm{Sh}\left( \rhow(\hatrho;q_1,p_1) \right),
\end{eqnarray}
which I call the ``Wigner-Shannon EE''.

In the one-body density matrix $\hatrho$ and its Wigner transformation,
quantum decoherence occurs and produces the EEs 
because of the factor $\langle \R';b_2|\R;b_2\rangle =\exp[-\frac{u}{4b^2} (\R-\R')^2]$, which 
originates in the reduction of the DOF of \pb.
Indeed, in the case of $u=0$, without this factor,  
 $\hat\rho^{(1)}=\{\hat\rho^{(1)}\}^2$ and the 
R\'enyi-2 and von Neumann EEs are zero
that corresponds to a pure single-particle state. 

Let us consider a semi-classical approximation of $\hatrho$.
The factor $\exp\left[-\frac{u}{4b^2}(\R-\R')^2\right]$, 
which is the source of the quantum decoherence,  
has a sharp peak around $\R'\approx \R$ with a width  $2b/\sqrt{u}$. 
I assume that the function $F(\R)$ is a slowly varying function compared with 
$\exp\left[-\frac{u}{4b^2}(\R-\R')^2\right]$, and it can be approximated as $F(\R')\approx F(\R)$. Then 
I obtain a semi-classical approximation,
\begin{eqnarray}
&&\frho(q_1,q_1')\clapprox  \frhocl(q_1,q_1')\nonumber\\
&&=\int d\R
\frac{ |f(\R)|^2 }{(\bs^2\pi)^{1/2}}\textrm{e}^{-\frac{1}{2\bs^2}(q_1-\R)^2 
-\frac{1}{2\bs^2}(q_1'-\R)^2 }\label{eq:cl-q},
\end{eqnarray}
where $f(\R)\propto F(\R)$ whose normalization is determined by 
$\int dq_1\frhocl(q_1,q_1) =\int d\R |f(\R)|^2 =1$. 
This corresponds to
\begin{eqnarray}
\hatrho&\clapprox &\hatrhocl\equiv
\int d\R |f(\R)|^2  |\R;\bs\rangle \langle \R;\bs|,\label{eq:semi-cl}
\end{eqnarray}
where the parameter $\R$ and $|f(\R)|^2$ are regarded as a
classical coordinate and a classical distribution of the 
composite particle, respectively. 
In the large $u$ limit---that is, the large $\pb$  mass limit---
$\rho(q_1,q_1')\to \rho^{\textrm{cl}}(q_1,q_1')$
and $\bs\to b$. Note that, $|\R \rangle$ and $|\R' \rangle $ are not
orthogonal to each other because of the quantum fluctuations 
of the $\pa$ position in the composite particle. 

In the semi-classical approximation given by Eq.~\eqref{eq:semi-cl}, 
the Wigner function is approximated as
\begin{equation}
\rhow(\hatrho;q_1,p_1)\clapprox2 \int d\R |f(\R)|^2 
\textrm{e}^{-\frac{1}{\bs^2}(\R-q_1)^2-\frac{\bs^2}{\hbar^2}p_1^2},
\end{equation}
which is nonnegative definite. Using $\hatrhocl$
I also define the EEs 
in the semi-classical approximation:
$\SRcl=\SR(\hatrhocl)$ and 
$S^\textrm{W-Sh,cl}=S^\textrm{W-Sh}(\hatrhocl)$.

For simple examples, I first 
consider the zero-momentum state of the composite particle 
in a finite volume $V$ described by a constant $F(\R)$. 
I assume that $V\gg b$ and 
the contribution of the box boundary can be ignored and obtain 
\begin{eqnarray}
\frho(q_1,q_1')&=&\frac{1}{V}
\textrm{e}^{-\frac{1}{4\bs^2}(q_1-q_1')^2},\\
\rhow(\hatrho;q_1,p_1)&=&\frac{2\bs\pi^{1/2}}{V}
\textrm{e}^{-\frac{\bs^2}{\hbar^2}p_1^2}.
\end{eqnarray}
In this case, $\frho(q_1,q_1')=\frhocl(q_1,q_1')$ is satisfied.
The R\'enyi-2, von Neumann, and Wigner-Shannon EEs are 
\begin{eqnarray}
\SR
&=&\loge V_{\rm eff}-\frac{1}{2}\loge(2\pi),\\
S^{\textrm{vN}}
&=&S^\textrm{W-Sh}= \SR+\frac{1}{2}(1-\loge 2),
\end{eqnarray}
where $V_{\rm eff}=V/\bs$ denotes the effective Volume size for the cm motion. 
These results are not valid for a small $V_{\rm eff}$ because of the box
boundary.

The one-body density matrix is diagonalized 
in the momentum space with a Gaussian distribution,
$\exp[-\frac{\bs^2}{\hbar^2}p_1^2]$. This
indicates that the one-body density matrix of a free composite particle is equivalent
to a thermal state of a single particle at finite temperature 
$kT= \hbar^2/2m \bs^2$. The temperature is of the same order as the 
mean kinetic energy, $\hbar^2/2m b^2$, of constituent particles 
confined in the composite particle. 
Strictly speaking EEs are not thermodynamic entropies, however, 
by associating the one-body density matrix of the free composite particle 
with a quantum mixed state of a single particle, I can propose an
interpretation of the entropy production and thermalization as follows: 
when the DOF of  $\pb$  are reduced, 
the quantum decoherence occurs, producing the entropy, and simultaneously,  
the intrinsic kinetic energy of the composite particle 
is converted into heat.

\begin{figure}[htb]
\begin{center}
	\includegraphics[width=7.5cm]{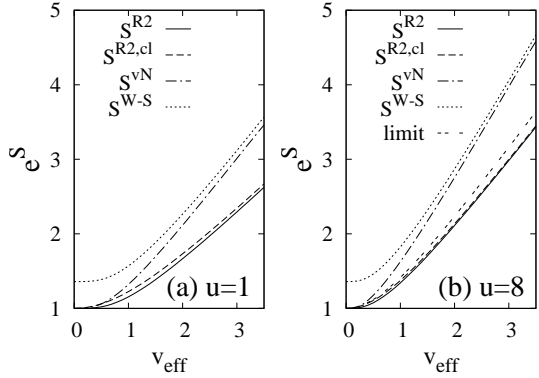} 	
\end{center}
  \caption{EE of a composite particle 
with a Gaussian distribution for (a) $u=1$ and (b) $u=8$. 
$e^S$ for the R\'enyi-2 ($\SR$ and $\SRcl$),
 von Neumann ($S^{\textrm{vN}}$), and Wigner-Shannon $S^{\textrm{W-Sh}}$) 
EEs are plotted as functions of the effective volume size, $v_{\rm eff}$.
The R\'enyi-2 EE in the large $u$ limit is also shown in the panel (b). 
\label{fig:s2}}
\end{figure}

Next, I consider a composite particle moving in an external ho potential, where
the lowest state of the composite particle is given by the Gaussian distribution,
$F(\R)=\textrm{e}^{-\frac{\R^2}{2B^2}}/(B^2\pi)^{1/4}$.
This gives the exact solution to the two-body wave function,
$\langle r_1,r_2|\Psi^{(2)}\rangle=\langle \rG,r|\Psi^{(2)}\rangle$, for 
ho potentials $U_{\rm ho}(M,\beta;\rG)+U_{\rm ho}(\mu,\bs;r)$,
with $M=(u+1)m$ and $\beta=\sqrt{B^2+u_1b^2}$.
In the $B=0$ limit, 
$\Psi^{(2)}$ describes a localized composite particle that corresponds 
to a non-entangled (uncorrelated) state of two constituent 
particles and has zero Renyi-2 and von Neumann EEs.
As $B$ enlarges and the delocalization of the cm of the composite particle 
grows, the EEs increase.
The Wigner function and EEs for $\hatrho$ are
\begin{eqnarray}
\rhow(\hatrho;q_1,p_1)&=&\frac{2\sqrt{\gamma} b}{\sqrt{b^2+B^2}}
\textrm{e}^{-\frac{1}{b^2+B^2}q_1^2-\frac{\gamma b^2}{\hbar^2}p_1^2},\\
\SR&=&\frac{1}{2}\loge (1+v_{\rm eff}^2)-\frac{1}{2}\loge \gamma,\\
S^\textrm{W-Sh}&=&
\SR+1-\loge 2,
\end{eqnarray}
where $\gamma=(1+(u+1)v_{\rm eff}^2)/(1+uv_{\rm eff}^2)$, and 
$v_{\rm eff}=B/b$ denotes the effective volume size.
The EEs increase as $v_{\rm eff}$ enlarges and
approaches $\loge v_{\rm eff}$ in the large 
$v_{\rm eff}$ limit. 
In the semi-classical approximation, the Wigner function 
and EEs are 
\begin{eqnarray}
\rhow(\hatrho;q_1,p_1)&\clapprox&
\frac{2\bs}{\sqrt{\bs^2+B^2}}
\textrm{e}^{-\frac{1}{\bs^2+B^2}q_1^2-\frac{\bs^2}{\hbar^2}p_1^2},\\
\SR&\clapprox&\SRcl=\frac{1}{2}\loge (1+v^2_{c,{\rm eff}}), \\
S^\textrm{W-Sh}&\clapprox&S^\textrm{W-Sh,cl}=
\SRcl+1-\loge 2,
\end{eqnarray}
where $v_{\rm c,eff}=B/\bs$. 
I show the EEs for $u=1$ and $u=8$ in Fig.~\ref{fig:s2}. $S^\textrm{vN}$
is calculated numerically, as was done in the previous paper \cite{enyo-ee1}.
$\SRcl$ for the semi-classical approximation agrees well  with
$\SR$ in the $v_{\rm eff}\ge 2$ case to within 10\% error for $u=1$, and the agreement is 
better for the larger mass ratio, $u=8$. $S^\textrm{W-Sh}$ has a constant shift $1-\loge 2$
(a constant scaling $\textrm{e}/2$ in the $\textrm{e}^S$ plot in Fig.~\ref{fig:s2})
from $\SR$, and it is finite even at
$v_{\rm eff}=0$. $S^\textrm{vN}$
starts from zero at $v_{\rm eff}=0$ and approaches $S^\textrm{W-Sh}$ as 
$v_{\rm eff}$ increases. 
As the mass ratio $u$ increases, the EEs converge on values in the large $u$ limit.

Finally, I connect the quantum decoherence in the one-body density matrix 
of the composite particle to coarse graining in the phase space distribution function of a single particle
and associate it with the Husimi function. 
Let us start from the Wigner transformation of the full two-body density matrix
$\hatrhotwo$,
\begin{eqnarray}
&&\rhow(\hatrhotwo;q_1,p_1,q_2,p_2)\nonumber\\
&&=\int d\eta d\xi
\langle q_1+\frac{\eta}{2},q_2+\frac{\xi}{2} |\hatrhotwo| q_1-\frac{\eta}{2},q_2+\frac{\xi}{2} \rangle
\nonumber\\ 
&&\times \textrm{e}^{-\frac{i}{\hbar}p_1\eta-\frac{i}{\hbar}p_2\xi}.
\end{eqnarray}
It is rewritten by a separable form in the phase space for the cm and relative coordinates as
\begin{equation}
\rhow(\hatrhotwo;q_1,p_1,q_2,p_2)=\rhow(\hat\rho^{(G)}_{\Phi_G};Q,P)
\rhow(\hat\rho^{(r)}_{\phi_\textrm{int}};q,p), \label{eq:two-body-cm-int}
\end{equation}
where $\hat\rho^{(G)}_{\Phi_G}=|\Phi_G\rangle\langle \Phi_G|$ and 
$\hat\rho^{(r)}_{\phi_\textrm{int}}=|\phi_\textrm{int}\rangle\langle \phi_\textrm{int}|$.
The Wigner function of the one-body density matrix can be written as
\begin{eqnarray}
&&\rhow(\hatrho;q_1,p_1)=\int \frac{dq_2dp_2}{2\pi\hbar} \rhow(\hatrhotwo;q_1,p_1,q_2,p_2)\nonumber\\
&&=\frac{1}{u_2} \int \frac{dQdP}{2\pi\hbar}\rhow(\hat\rho^{(G)}_{\Phi_G};Q,P) \nonumber\\
&&\times \rhow \left(\hat\rho^{(r)}_{\phi_\textrm{int}};\frac{q_1-Q}{u_2},p_1-u_1P\right)\label{eq:cg-general}\\
&&=\frac{1}{u_2} \int \frac{dQdP}{\pi\hbar}\rhow(\hat\rho^{(G)}_{\Phi_G};Q,P) \nonumber\\
&&\times \textrm{e}^{-\frac{1}{\bs^2u_2^2} (q_1-Q)^2-\frac{\bs^2}{\hbar^2}(p_1-u_1P)^2}.
\label{eq:coarse-graining}
\end{eqnarray}
Here I use relations $q=(q_1-Q)/u_2$, $p=p_1-u_1P$, and the transformation
$dq_2 dp_2 = |J| dQdP$ with the determinant of Jacobian $|J|=1/u_2$. This means that
$\rhow(\hatrho;q_1,p_1)$ is regarded as a coarse grained distribution function of 
$\rhow(\hat\rho^{(G)}_{\Phi_G};Q,P)$ with a Gaussian smearing. In other words, 
the quantum decoherence caused by the reduction of the DOF of $\pb$  can be 
interpreted as the coarse graining in the phase space distribution of a single-particle state.
It is important that, if the internal DOF are decoupled from the cm motion of the composite particle,
Eq.~\eqref{eq:cg-general} describes a general form of the coarse grained distribution function 
that corresponds to the Wigner function of $\hatrho$. 

Let us consider the $u=1$ case and associate the coarse graining in Eq.~\eqref{eq:coarse-graining} 
with the Husimi function. Eq.~\eqref{eq:coarse-graining} for $u=1$ is rewritten 
as
\begin{eqnarray}
&&\rhow(\hatrho;q_1,p_1)=2\rhohh(\hat\rho^{(G)}_{\Phi_G};b_G^2;q_1,p_1),\\
&&\rhohh(\hat\rho^{(1)};\alpha;q,p)\nonumber\\
&&\equiv \int \frac{dQd P}{\pi(\hbar/2)} \int d\eta 
\langle Q+\frac{\eta}{2}|\hat\rho^{(1)}|Q-\frac{\eta}{2}\rangle 
\textrm{e}^{-\frac{i P\eta}{\hbar/2}} \nonumber\\
&&\times \textrm{e}^{-\frac{1}{\alpha}(q-Q)^2-\frac{\alpha}{(\hbar/2)^2}(p-P)^2},
\label{eq:fusimi-half}
\end{eqnarray}
where $\rhohh$ is normalized as $\int \frac{dqdp}{2\pi(\hbar/2)}\rhohh=1$.
$\rhohh$
eventually has the same form as 
the normal Husimi function, except for scaling of Planck's constant 
$\hbar\to \hbar/2$. I call $\rhohh$ ``$\hbar/2$-Husimi function''.
It is clear that $\rhow(\hatrho;q_1,p_1)$ for the two-body state $\Psi^{(2)}$
is equivalent to twice the $\hbar/2$-Husimi  function for 
the single-particle state, $|\Phi_G\rangle$.
The Gaussian smearing in the coarse graining originates from the reduction of the DOF of $\pb$
in $\rhow(\hat\rho^{(r)}_{\phi_\textrm{int}};(q,p))$ as shown previously.
Note that the $\hbar/2$-Husimi  function is not a distribution function for a physical single-particle state, 
but is regarded as a ``distribution'' function defined in the down-scaled phase space, $\hbar\to \hbar/2$.
The reason for the down scaling $\hbar\to \hbar/2$ is that the $(q_2,p_2)$ phase space is 
scaled down in the transformation from $(q_1,p_1,q_2,p_2)$ to  $(q_1,p_1,Q,P)$.

Considering the one-to-one correspondence between 
$\rhow(\hatrho;q_1,p_1)$ and $\rhohh(\hat\rho^{(G)}_{\Phi_G};q_1,p_1)$, I can connect
entropies defined by the $\hbar/2$-Husimi function to EEs as
\begin{eqnarray}
&&S^{\textrm{Wehrl}}_{\hbar/2}(\hatrhoG) =
S^{\textrm{W-Sh}}(\hatrhotwo)+\loge 2,\\
&&S^{\textrm{R2-Wehrl}}_{\hbar/2}(\hatrhoG)=
\SR(\hatrhotwo)+\loge 2.
\end{eqnarray}
Here I define the Wehrl entropy and the R\'enyi-Wehrl entropy of the order 2
in the down-scaled phase space as
\begin{eqnarray}
&&S^{\textrm{Wehrl}}_{\hbar/2}(\hat\rho^{(1)})=-  \int \frac{ dqdp }{2\pi(\hbar/2)} \rhohh(\hat\rho^{(1)};q,p)\loge \rhohh(q,p),\nonumber\\
&&S^{\textrm{R2-Wehrl}}_{\hbar/2}(\hat\rho^{(1)})=-\loge\left[ \int\frac{ dqdp }{2\pi(\hbar/2)} \{\rhohh(\hat\rho^{(1)};q,p)\}^2\right]
\nonumber.
\end{eqnarray}

In summary, I investigate the quantum decoherence in the one-body density matrix of the composite particle that comprised 
two correlated particles in the inert composite particle approximation. 
Because of the two-body correlation in the composite particle, the quantum decoherence occurs by the reduction of 
the DOF of the second particle. 
As the delocalization of the distribution of the composite particle grows, the entanglement entropy increases.
I found a one-to-one correspondence between 
the quantum decoherence in the reduced density matrix and the coarse graining in the phase space
distribution, which is related to the Husimi-like function defined in the down-scaled phase space.
In the present paper, the inert composite particle approximation is applied to static systems but it can also be extended to time-dependent systems if the energy scale of the internal DOF of the composite particle is decoupled from that of the external DOF.
The present study may shed light on the fundamental problems of quantum decoherence 
and coarse graining which produces entropies in quantum systems.

This work was supported by 
JSPS KAKENHI Grant No. 26400270.

\end{document}